\documentclass[thmsb,11pt]{article}
\usepackage{amsfonts,amsmath,amsthm, amstext,amssymb}
\usepackage{mathpazo}
\usepackage{fullpage}
\usepackage{color}
\usepackage{graphicx}
\usepackage{enumerate}
\usepackage{pgf, tikz}
\usepackage{epic}
\usepackage{multirow}
\usepackage{afterpage}
\usepackage{lscape}
\usepackage{lineno}
\usepackage{cases}
\usepackage{array}
\usepackage[round,authoryear]{natbib}
\usepackage[english]{babel}
\usepackage{makeidx}
\usepackage[compatibility=false]{caption}
\usepackage{subcaption}
\usepackage{mathpazo} 
\usepackage{multimedia}

\usepackage[round,authoryear]{natbib}
\newtheorem{proposition}{Proposition}
\newtheorem{assumption}{Assumption}

\begin{document}

\title{{\Large The Dynamics of Leverage and the Belief Distribution of Wealth\thanks{Dedicated to the memory of Peter Flaschel, with gratitude for his mentorship and kindness. We thank Mausumi Das, E. Somanathan, two anonymous referees, and seminar participants at the Delhi School of Economics for helpful feedback on an earlier version.}}}
\author{Bikramaditya Datta\thanks{Department of Economic Sciences, Indian Institute of Technology, Kanpur} \and Rajiv Sethi\thanks{%
Department of Economics, Barnard College, Columbia University and the Santa
Fe Institute.}}
\maketitle

\begin{abstract}
\noindent The scale and terms of aggregate borrowing in an economy depend on the manner in which wealth is distributed across potential creditors with heterogeneous beliefs about the future. This distribution evolves over time as uncertainty is resolved, in favor of optimists if loans are repaid in full, and in favor of pessimists if there is widespread default. We model this process in an economy with two assets---risky bonds and risk-free cash. Within periods, given the inherited distribution of wealth across belief types, the scale and terms of borrowing are endogenously determined. Following good states, aggregate borrowing and the face value of debt both rise, and the interest rate falls. In the absence of noise, wealth converges to beliefs that differ systematically from the objective probability governing state realizations, with greater risk-aversion associated with greater optimism. In the presence of noise, the economy exhibits periods of high performance, punctuated by periods of crisis and stagnation. 

\end{abstract}

\thispagestyle{empty} \newpage

\section{Introduction}

Debt plays multiple interlocking roles in a market economy---it finances expenditure, increases vulnerability to negative shocks, expands the universe of assets available to investors, and results in changes over time in the distribution of wealth. Among creditors, those who are most optimistic about future conditions shoulder more risk, and  prosper relative to more pessimistic individuals if their optimism turns out to have been warranted. The resulting change in the distribution of wealth, towards those with more optimistic beliefs, affects the volume and terms of lending. It can do so in ways that are self-reinforcing, amplifying and prolonging a boom. By the same token, a deterioration of economic conditions shifts the distribution of wealth towards those who were more pessimistic to begin with, and thus chose to invest safely. This process, too, can be self-reinforcing, and can result in a period of stagnation or economic under-performance. 

 In this paper, we explore the manner in which the scale and terms of lending evolve over time in relation to the manner in which wealth is distributed across individuals with varying levels of optimism. We call this the \textit{belief distribution of wealth}. Our starting point is the assumption that individuals have heterogeneous prior beliefs, and thus agree to to disagree about future economic conditions. This leads to heterogeneity in portfolio choices even when risk preferences are homogeneous. The belief distribution of wealth in any given period determines the total volume of lending that can be sustained, and the price of debt. This, in turn, affects the likelihood of positive and negative shocks. Once these shocks are realized, the belief distribution of wealth shifts---towards optimists if loans are repaid in full and towards pessimists if there is default. The movement of the economy through time can thus be viewed as the stochastic evolution of the belief distribution of wealth, with the scale and terms of lending and the probability of default all determined endogenously. 
 
 In any given period, given the inherited belief distribution of wealth, the set of investors can be partitioned into three groups---optimists who invest only in risky bonds, pessimists who choose only risk-free cash, and intermediate belief types who choose combinations of both. Uncertainty may be resolved in one of two ways: in a good state all loans are repaid in full, while there is partial default in the bad state. We show that following a good state, the volume of lending increases---the amount borrowed and the amount to be repaid at maturity both rise. The bond price also rises, so the interest rate falls, and (conditional on beliefs) portfolios shift towards smaller allocations in bonds.  A bad shock results in precisely the opposite effects: the volume of lending contracts, bond prices fall, and (conditional on beliefs) portfolios shift towards larger allocations in bonds. Through this process, the portfolio choices of investors, the belief distribution of wealth, the scale and terms of borrowing, and the likelihood of default evolve over time.
 
This system exhibits some interesting long run properties. When preferences are risk neutral (or more generally insufficiently risk averse), the long run distribution of wealth comes to be concentrated on beliefs that are excessively pessimistic relative to the objective probability of a good state. The opposite happens when preferences are highly risk averse (with a coefficient of relative risk aversion exceeding unity). In this case convergence occurs to beliefs that are more optimistic than the objective probability of a good state. The intuition for these perhaps surprising findings is that the maximization of wealth in the long run requires individuals with accurate beliefs to invest as if they had log utility. When preferences depart from this, excessive risk-aversion can counteracted by overoptimism, while excessive risk-taking can be balanced by undue pessimism.

When the objective probability of default itself depends on the scale and terms of borrowing, and there is some noise in the process, the economy exhibits periods of high lending volume, with low interest rates and low default probabilities, but these are punctuated by periods of crisis and stagnation. 

 A key distinction in our analysis is between belief \textit{accuracy} and belief \textit{profitability}. As noted by \citet{de1990noise}, departures from belief accuracy can be lucrative in the long run if they induce people to hold more risk, since risk is rewarded in financial markets. In our case, whether excessive optimism or excessive pessimism is more lucrative than accuracy depends on preferences in relation to those that would maximize the growth of wealth. In particular, long run wealth maximization occurs when investors with accurate beliefs maximize the expected value of log wealth \citep{kelly1956rule}. When preferences depart from this, undue pessimism can compensate for excessive risk taking, while excessive optimism can compensate for insufficient appetite for risk. 

In our model, profitable beliefs (even if inaccurate) become more prevalent in the population over time, as wealth accrues disproportionately to those who hold such beliefs. This mechanism has been previously explored in \citet{sethi1996endogenous}, and is formally equivalent to models of learning in which individuals adopt forecasting rules that have proved to be lucrative in the past \citep{brock1997rational}. This may be contrasted with learning from data, by comparing expectations to realizations, as in the literature descended from \citet{blume1982learning}, \citet{marcet1989convergence} and \citet{evans2001learning}.\footnote{See \citet{alchian1950}, \citet{friedman1953}, \citet{cootner1964}, and \citet{fama1965} for various versions of the idea that investors with incorrect beliefs will lose money relative to those with more accurate assessments, and will eventually disappear under pressure of competition. \citet{sandroni2000}, \citet{blume2006}, \citet{kets2014betting}, and \citet{dindo2020wisdom} identify conditions under which this market selection hypothesis fails to hold.}  In our model such learning from data would lead agents to adopt beliefs that are more accurate, but that erode their wealth in the long run.

An alternative approach to learning is based on mimetic contagion, as in \citet{weidlich1983concepts}, \citet{lux1995herd}, \citet{franke2014aggregate}, and \citet{flaschel2018macroeconomic}. While we do not consider this learning mechanism, we conjecture that incorporating it into the model would affect the speed of convergence and the likelihood of transitions between high and low leverage regimes, without materially affecting the qualitative properties of our model.

The emphasis on credit as a driving force in economic fluctuations has a long history in economics, dating back at least to \citet{hawtrey1919currency} and \citet{lavington1925trade}. The role of leverage in fueling expansions, and setting the stage for periodic crises was a central theme in the writings of Hyman \citet{minsky1975john,minsky1982can,minsky1986stab}, who viewed his work as an extension of ideas in \citet{keynes1936the}.\footnote{There have been a number of attempts to model the dynamics implicit in Minsky's relatively informal treatment of the topic, including \citet{guttentag1984credit}, \citet{taylor1985minsky}, \citet{sethi1992dynamics,sethi1995evolutionary}, \citet{skott1994modelling}, \citet{semmler1987macroeconomic}, \citet{chiarella2000disequilibrium}, \citet{gatti2005new}, \citet{charpe2009overconsumption}, \citet{eggertsson2012debt}, \citet{keen2013monetary}, and \citet{bhattacharya2015reconsideration}. See \citet{nikolaidi2017minsky} for a survey.} Following the global financial crisis of 2008-09 there has been renewed interest in the role of these mechanisms  \citep{admati2013bankers}. Some of the resulting literature has adopted a framework with heterogeneous prior beliefs \citep{geanakoplos2010leverage,fostel2008leverage}.\footnote{The value of allowing for departures from the common prior assumption, especially in studies of asset prices and trading, was demonstrated by \citet{harrison1978speculative}. The measurement and modeling of expectations is an active area of research; see \citet{d2023data} for a survey in which the authors document biases, cross-sectional dispersion, and volatility over time in beliefs.}  Our approach here builds directly on \citet{che2014credit}, with the exception that the total amount of borrowing in any period is taken to be the maximum that can be funded, given the belief distribution of wealth, instead of being exogenously given. This model is embedded in a sequence of periods, so that the evolution of leverage and the wealth distribution can be examined. As in \citet{geanakoplos2010leverage}, the model exhibits high leverage and elevated asset prices in booms, and low leverage and depressed asset prices in times of crisis.

There are two distinct visions of the manner in which capitalist economies generate periodic crises. One perspective, descended from Wicksell, Frisch, and Stutsky, views the economy as being fundamentally stable but buffeted by unpredictable shocks that result in damped propagation effects. The other views the economy as being intrinsically unstable, even in the absence of shocks, with nonlinear effects keeping trajectories bounded. \citet{zarnowitz1985recent} categorizes the former as exogenous theories, and the latter as endogenous.\footnote{\citet{gabisch2013business} refer to the former as shock-dependent theories and the latter as shock-independent. Shock-dependent theories currently dominate the research mainstream, though shock-independent theories have a long history \citep{kaldor1940model,hicks1950contribution,goodwin1951nonlinear,tobin1975keynesian,grandmont1985endogenous,foley1987liquidity}. See \citet{chiarella2012reconstructing,chiarella2013reconstructing,chiarella2015reconstructing} for a comprehensive overview and further development of the shock-independent approach.} The model developed here, which is methodologically close to \citet{aoki2007reconstructing}, does not fit neatly into either one of these categories. Shocks matter in inducing transitions between high and low leverage regimes, but positive feedback effects result in potentially large and sustained departures from steady growth.

\section{The Model}

\subsection{Preliminaries}

Consider a infinite sequence of periods indexed by $t=1,2,....$ Let
$B_{t}$ denote the amount of aggregate borrowing in period $t$.
We consider only one period loans, implemented by the issue of $Q_{t}$
bonds, each with unit face value and price $p_{t}$. The borrowing results in a random aggregate revenue $Y_t$ dependent on the realization of an aggregate shock, which can either be positive or negative. This revenue is recoverable by lenders if loans are not repaid in full. 

The scale and terms of borrowing in any period, as well as the two possible values of realized revenue, are all endogenous and depend on the distribution of wealth among potential lenders. 
There are a set of investors who have heterogeneous beliefs about the probability distribution governing the aggregate
uncertainty. An investor of type $\theta$ believes that the likelihood
of a positive shock is $\theta$. The set of possible beliefs is finite and given by $\Theta=\{ \theta_{1},\theta_{2}...,\theta_{n} \}$, 
where $0 = \theta_1 < \theta_2 < ... < \theta_n \ =1$. Note that $\Theta$ is simply the set of possible beliefs; as we show below, the share of wealth held by those with very extreme beliefs (relative to the objective probability distribution) will approach zero endogenously over time.

A key feature of the model is the distribution of wealth across belief types, which we call the \textit{belief distribution of wealth}. Let $W_{t}(\theta_i)$ denote
the wealth of investor of type $\theta_i$ at time $t.$ The aggregate wealth held by investors at time $t$ is then 
$$\bar{W}_{t} \equiv \sum_{\theta_i \in \Theta}W_{t}(\theta_i).$$
Let $f_t(\theta_i) = W_{t}(\theta_i)/{\bar{W}_{t}}$ denote the share of total wealth held by belief type $\theta_i$, so the distribution function for wealth is $$F_{t}(x) = \sum_{\theta_i \leq x} f_t(\theta_i).$$
This distribution evolves over time in ways that depend on portfolio choices, bond prices, and shock realizations, all of which are endogenous.

It is convenient to normalize variables by total wealth in each period. Accordingly, define borrowing per unit wealth as $b_{t} \equiv {B_{t}}/{\bar{W}_{t}}$, and the bond issue per unit wealth as $q_{t} \equiv {Q_{t}}/{\bar{W}_{t}}$. Let $y_t = Y_t/\bar{W}_t$ denote normalized revenue in period $t$, which depends on the shock realization, as well as the level of (normalized) borrowing $b_t$, in a manner specified below. 

Given the wealth distribution $f_t(\theta)$ at the start of a period, the variables $b_t$, $q_t$, and $p_t$ are endogenously determined. This will affect the probability distribution over good and bad states, and the distribution of $y_t$. The realized value of $y_t$ then determines the distribution of wealth $f_{t+1}(\theta)$ in the subsequent period, and the process repeats. The evolution of the economy may therefore be described by a stochastic sequence of wealth distributions $\{ f_t(\theta) \}_{t=1}^\infty$, with all other variables endogenously determined. 

\subsection{The Scale and Terms of Borrowing}

There are two possible revenue outcomes depending on the realization of the aggregate shock: a high value $h(b_{t})$ 
and a low value $l(b_{t})$.
We assume that a larger quantity of borrowing (relative to total wealth) results in the choice of progressively inferior projects that result in lower payoffs even if successful. That is, although revenue rises with total borrowing in either state, it does so at a decreasing rate.\footnote{This assumption ensures uniqueness of equilibrium within periods, and allows us to focus on the evolution of the system over time. While the assumption is fairly standard in the literature on aggregate investment, dating back at least to \citet{keynes1936the} on the marginal efficiency of capital, it rules out increasing returns through network effects and other sources. We do consider positive feedback effects of leverage on the probability of good states in Section \ref{s:endogenous}.} Specifically, we assume that $h$ and $l$ are concave, with $h$ being strictly concave (we allow for the possibility that $l$ is linear). In addition, these functions satisfy $l(b_t) \le b_t \le h(b_t)$, with strict inequality if $b_t \in (0,1)$.

The scale of borrowing is bounded above by the constraint that there must be enough revenue in the good state to pay back all loans in full. We shall assume that this constraint is binding:
\begin{equation} \label{issue}
    q_t = h(b_t).
\end{equation}
The rationale is that projects yield significant intangible and non-pledgeable payoffs to borrowers, which induces them to implement projects as long as they can get them funded. Hence the amount repaid in
the good state exhausts all available revenue, and there must be (partial) default in the bad state. The bond price must satisfy 
\begin{equation} \label{price}
    p_t q_t = b_t 
\end{equation}
in order that that borrowing requirement is met. 

In order to complete the model, we need to specify the behavior of creditors. Let $\sigma_{it}$ denote the share of wealth invested in bonds by an investor with beliefs $\theta_i$, given the conditions prevailing in period $t$. This will depend, of course, on the period $t$ bond price, but also on the total amount of borrowing, since this affects recovery in the case of default. Conditional on a positive shock, an investor with portfolio share $\sigma_{it}$ on bonds will obtain 
$$ 1- \sigma_{it} + \frac{\sigma_{it}}{p_t}$$  per unit of wealth invested, since she will have purchased $1/p_t$ bonds with unit face value that are repaid in full, and held the rest of her wealth in cash that is risk free and interest free. Conditional on a negative shock, each bond pays $\l(b_t)/q_t$, so the corresponding payoff per unit of wealth invested will be 
$$ 1-\sigma_{it} + \frac{\sigma_{it}}{p_t} \frac{l(b_t)}{q_t}.$$ 
We shall assume that creditors choose portfolios to maximize expected utility, based on preferences that satisfy constant relative risk aversion. That is, they choose $\sigma_{it}$ to maximize 
\begin{equation} \label{e:objective}
    \theta_i u \left( 1- \sigma_{it} + \frac{\sigma_{it}}{p_t} \right) + (1-\theta_i) u \left( 1-\sigma_{it} + \frac{\sigma_{it}}{p_t} \frac{l(b_t)}{q_t} \right),
\end{equation}
where 
\begin{equation*}
u(w) = \frac{w^{1-\gamma}}{1-\gamma}\label{util}
\end{equation*}
for some risk-aversion parameter $\gamma \neq 1$, and  $u(w) = \log(w)$ when $\gamma = 1$. Since portfolio choice with such preferences does not depend on wealth, all investors with belief $\theta_i$ will choose the same share $\sigma_{it}$ in bonds in any given period. This allows us to focus on the distribution of wealth across belief types without concern for how wealth is distributed \textit{among} individuals holding the same belief. The final equilibrium condition states that the total amount of borrowing must equal the total amount that investors wish to lend:
\begin{equation} \label{market_clear}
    b_{t} = \sum_{\theta_i \in \Theta} \sigma_{it} f_{t}(\theta_i)
\end{equation}

The last ingredient in the model is the objective probability distribution governing shock realizations, which we allow to depend on the scale of borrowing. Let $\pi(b_t)$ denote the probability of a positive shock when borrowing is $b_t$. Note that the system is self-referential, in the sense that the true distribution may depend on the manner in which wealth is distributed across subjective belief types. 

Given the wealth distribution $f_t(\theta)$ at the start of period $t$, conditions (\ref{issue}--\ref{market_clear}) together determine the values of $b_t$, $q_t$, $p_t$, and the portfolio shares $\sigma_{it}$. The scale of borrowing, in turn, determines the probability $\pi(b_t)$ of a positive shock. The shock realization leads to a change in the belief distribution of wealth, with a positive shock shifting this distribution towards optimists who invested in the bond, and a negative shock shifting it towards pessimists who chose safety. This determines the belief distribution of wealth in the next period, and the process repeats. 

\section{Equilibrium and Dynamics}

Within the class of CRRA preferences, risk-neutrality corresponds to $\gamma = 0$ and log utility to $\gamma =1$. We shall focus on these two cases analytically, but present numerical results also for other preference parameters. 

\subsection{Risk Neutrality} \label{s:riskneutral}

With risk-neutral preferences, in any given period $t$, all belief types will choose to invest entirely in bonds or entirely in cash, with the possible exception of a threshold belief type who may be indifferent between the two. Define $\bar{\theta}_{t}$ as the critical belief type such that
\begin{equation} \label{crit}
\bar{\theta}_{t} = \min\left\{ \theta \in \Theta \, \left| \, 1 - F_t(\theta) < b_t \right. \right\} .
\end{equation}
That is, the critical belief type is the most pessimistic investor who must be induced to lend if the demand for funds $b_t$ is to be met---the combined wealth of all those who are more optimistic than $\bar{\theta}$ is insufficient to meet the demand. In equilibrium, investors holding this belief will invest all or part of their wealth, while those more optimistic will invest all, and those more pessimistic will keep their wealth in cash. 

Assuming that cash earns no interest, the critical belief type will be induced to invest in bonds at time $t$ only if
\begin{equation*} 
\bar{\theta} + (1-\bar{\theta}) \frac{l(b_t)}{q_t} \geq  p_t.
\end{equation*}
This follows from the fact that $p_t$ buys a bond that pays one unit in the good state and $l(b_t)/{q_t}$ in the bad state, when there is (partial) default. Using (\ref{issue}--\ref{price}), the above equation may be written:
\begin{equation} \label{indiff}
\bar{\theta} \frac{h(b_t)}{b_t} + (1-\bar{\theta}) \frac{l(b_t)}{b_t} \geq  1.
\end{equation}
If the above inequality is strict, the investor will use her entire wealth to buy bonds. If the condition holds with equality, the critical belief type will be indifferent, and may keep part of her wealth in cash. 

Given the wealth distribution $f_t(\theta)$ at the start of period $t$, the scale and terms of borrowing are uniquely determined, as is the marginal belief type. 

\begin{proposition} \label{p:unique}
Given any belief distribution of wealth $f_t(\theta)$, the values of $\bar{\theta}_t$, $b_t$, $q_t$, and $p_t$ are uniquely determined.
\end{proposition}

The unique equilibrium level of borrowing $b_t$ determines the objective probability of a positive shock, and the realized value generates the subsequent belief distribution. A positive shock shifts the distribution of wealth in favour of those who have made loans, while a negative shock shifts the distribution towards those who have not. This follows from the fact that $l(b) < b < h(b)$ at any non-degenerate level of borrowing. A positive shock thus
implies that $F_{t+1}\left(\theta\right)\leq F_{t}\left(\theta\right)$,
while a negative shock implies that $F_{t+1}\left(\theta\right)\geq F_{t}\left(\theta\right)$. 

In addition, if there is any change in the volume of lending following a positive shock, this volume rises, the terms become less favorable to lenders, and a more optimistic belief type becomes marginal. The effects of a negative shock are analogous (operating in reverse): 

\begin{proposition} \label{p:compstat}
{If $y_{t}=h\left(b_{t}\right)$,
then $b_{t+1}\geq b_{t},\:q_{t+1}\geq q_{t},\:p_{t+1}\geq p_{t},$
and $\bar{\theta}_{t+1}\geq\bar{\theta}_{t}$.} {These
inequalities are reversed if} $y_{t}=l\left(b_{t}\right).$
\end{proposition}

A sequence of positive shocks causes the belief distribution of wealth to shift towards optimists, resulting in increased borrowing at more favorable terms to borrowers. That is, the interest rate on loans falls even as borrowing is rising.

\begin{figure}[t]
\begin{center}
\includegraphics[width=5in]{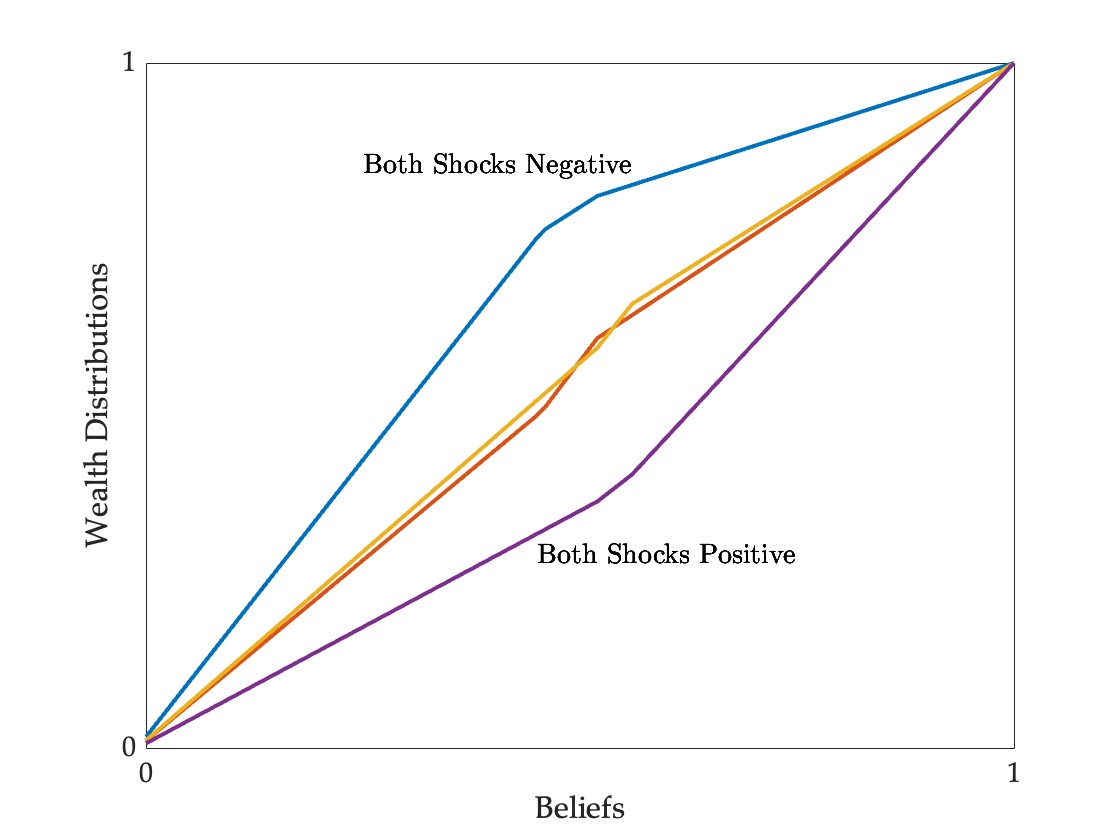}
\end{center}
\vspace{-0.25in}
\caption{Possible Wealth Distributions after Two Periods\label{dists2}}
\end{figure}

To illustrate, consider a  numerical example. Suppose that $h(b) = \sqrt b$ and $l(b) = b/2$. The set of possible beliefs is $\Theta = (0, 0.01, ..., 0.99, 1)$, and the initial belief distribution of wealth is uniform. Consider the first two periods, for which there are four possible paths. In the first period equilibrium we have $b = 0.4752$, $p = q = 0.6894$, and $\bar{\theta} = 0.53$. The optimists with beliefs $\theta \ge \bar{\theta}$ each buy 1.4506 bonds with unit face value, while the rest of the investor population is more pessimistic and remains in cash. After the good state, the optimists are paid in full. Total wealth rises to 1.2141, and the share held by the optimists rises to 69.48\%. After the bad state there is partial default, total wealth falls to 0.7624, and the share held by optimists falls to 44.40\%. 

After the second period, there are four possible wealth distributions, depending on the types of shocks that arise. These distributions are shown in Figure \ref{dists2}. When both states are bad, wealth moves towards pessimists, and when both are good the optimists prosper. But for the two intermediate cases the distribution functions intersect and the wealth distribution depends on the precise sequence of shocks.

Path dependence arises because the set of those who choose to invest in the second period is sensitive to the outcome in the first period. When the first period shock is positive, the critical belief type rises and some investors who invest and gain in the first period choose not to invest in the second. This is because the bond price rises---even though their belief has not changed, the were close to indifferent at the first period price and thus choose not to invest at the higher second period price. This set of investors then escapes the bad shock in the second period since they remain in cash. Had the sequence of shocks been reversed, investors at the margin would have lost in the initial period, then invested again in the second (since the bond price and the critical belief type both fall). They would thus be worse off than if the good shock had come first.

\subsection{Log Utility}

Now consider the case of $\gamma = 1$, with preferences are given by $u(w) =\log(w)\label{log-util}$. An investor with belief $\theta_i$ then chooses $\sigma_{it}$
to maximize 
\begin{equation}
    \theta_i \log \left( 1- \sigma_{it} + \frac{\sigma_{it}}{p_t} \right) + (1-\theta_i) \log \left( 1-\sigma_{it} + \frac{\sigma_{it}}{p_t} \frac{l(b_t)}{q_t} \right).
\end{equation}
Define $H_t$ and $L_t$ as follows:
$$H_t \equiv \frac{1}{p_{t}}-1, \qquad L_t \equiv1- \frac{1}{p_{t}} \frac{l(b_{t})}{q_{t}}.$$
Here $H_t$ is the gain per unit of wealth invested in bond conditional on a positive shock, and $L_t$ is the loss per unit of wealth invested in bonds conditional on a negative shock. Both are positive, and depend on the triple $(b_t,p_t,q_t)$. Define $\theta^{\min}_t$ and $\theta^{\max}_t$ as follows:
$$ \theta_t^{\min} \equiv \frac{L_t}{H_t+L_t}, \qquad \theta_t^{\max} \equiv \frac{H_t L_t+L_t}{H_t+L_t}.  $$ 
It may be verified that $0 < \theta_t^{\min} < \theta_t^{\max} < 1$ and that the optimal portfolio choices satisfy: 
\begin{equation} \label{e:logsolution}
    \sigma_{it} = \begin{cases}
0 & \textrm{if}\:\theta_i \leq \theta_t^{\min}\\
\frac{\theta_i H_t - (1-\theta_i)L_t}{H_t L_t} & \textrm{if}\: \theta_t^{\min} < \theta_i \leq \theta_t^{\max}\\ 
1 & \textrm{otherwise}
\end{cases},
\end{equation}
The uniqueness and comparative statics results obtained for the risk neutral case can be extended to the case of log utility case if the following holds:
\begin{assumption} \label{ass1}
The functions $h(b)$ and $l(b)$
satisfy $\lim_{b\downarrow0} {h(b)}/{b} = \infty$, $\lim_{b\downarrow0}l(b)/b < 1$, and $l(1) < 1 = h(1)$.
\end{assumption}
These conditions ensure that the level of borrowing in equilibrium is interior, and are sufficient to yield  uniqueness and unambiguous comparative statics for the case of log utilty. 

\begin{proposition} \label{p:unique-log} Suppose $\gamma=1$ and Assumption \ref{ass1} holds. Then, given any belief distribution of wealth $f_{t}(\theta)$, the values
of $b_{t}$, $q_{t}$, and $p_{t}$ are uniquely determined. {If $y_{t}=h(b_{t})$,
then $b_{t+1}\geq b_{t},\:q_{t+1}\geq q_{t},$ and $p_{t+1}\geq p_{t},$.}
{These inequalities are reversed if} $y_{t}=l(b_{t}).$
\end{proposition}

This reproduces for the log utility case the content of Propositions \ref{p:unique}--\ref{p:compstat}, which applied to the case of risk-neutrality. The importance of considering this case  will become apparent when we consider the long run dynamics in the next section. 

\section{Long Run Dynamics} \label{s:longrun}

Suppose for the moment that the objective probability distribution over states is exogenous, unaffected by the distribution of wealth and the volume of lending. That is, $\pi(b) = \pi^*$ for some $\pi^* \in (0,1)$. In this case one may ask whether the belief distribution of wealth converges to accuracy in the long run, in the sense that all wealth is concentrated at the belief type that has the correct or "rational" expectation. 

\begin{figure}[t]
\begin{center}
\includegraphics[width=5in]{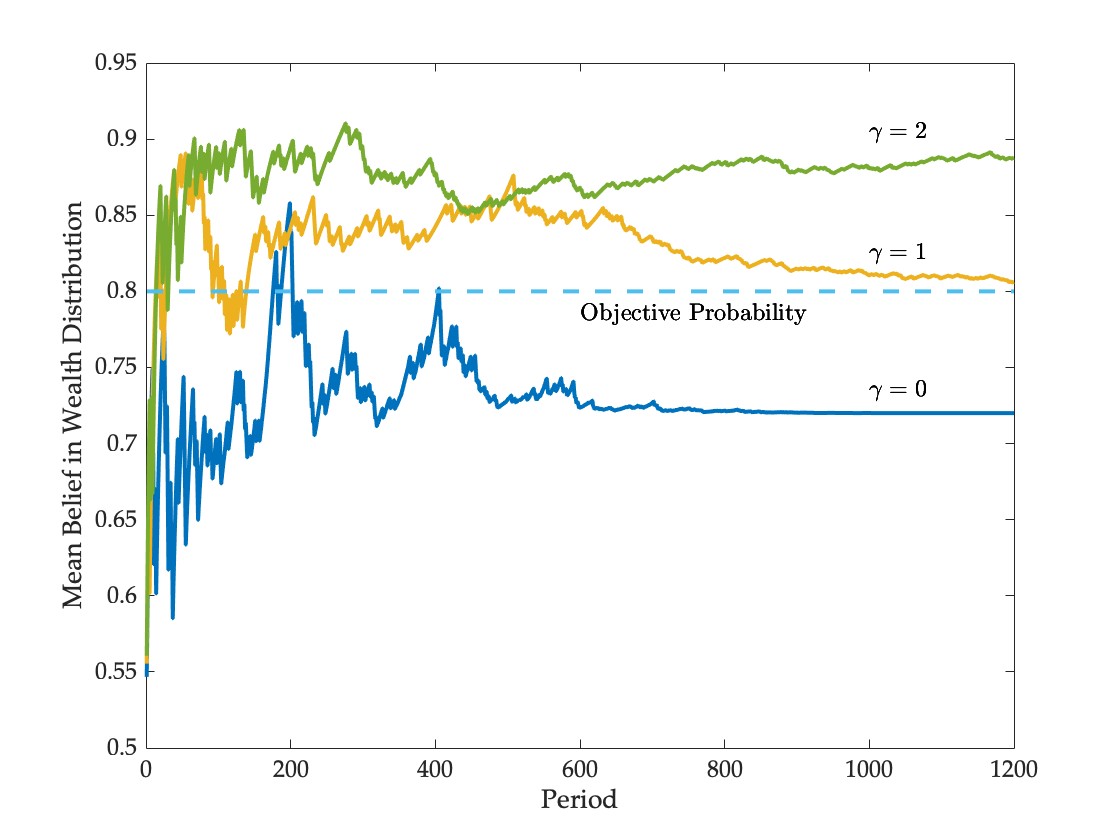}
\end{center}
\vspace{-0.25in}
\caption{Evolution of mean beliefs for different values of risk aversion\label{f:three_gammas}}
\end{figure}

It turns out that the distribution of wealth converges to beliefs that are systematically inaccurate in the long run, with a positive association between risk-aversion and optimism. This can be seen in Figure \ref{f:three_gammas}, which shows three sample paths, each with different risk preferences, but with the objective probability of a good state fixed exogenously at $\pi^* = 0.8$. Only in the borderline case $\gamma = 1$ (corresponding to log utility) do we get anything approximating convergence to accurate beliefs. With risk neutral preference the long run belief distribution is excessively pessimistic (relative to the objective probability of a good state). And when preferences are more risk averse than the log utility benchmark, the beliefs are excessively optimistic in the long run.

To understand this phenomenon, recall from \citet{kelly1956rule} that the investment strategy that results in the greatest wealth accumulation in the long run corresponds to log utility. Relative to this, risk-neutral individuals with accurate beliefs invest too heavily in bonds at any given price. Their wealth grows less slowly in the long run than the wealth of risk-neutral types whose beliefs are somewhat more pessimistic. The pessimism counteracts the excessive risk taking that is entailed by risk neutrality. For similar reasons, when risk aversion is too high relative to the log utility benchmark, optimism serves to counteract the reluctance to shoulder sufficient risk.

\begin{figure}[t]
\begin{center}
\includegraphics[width=5in]{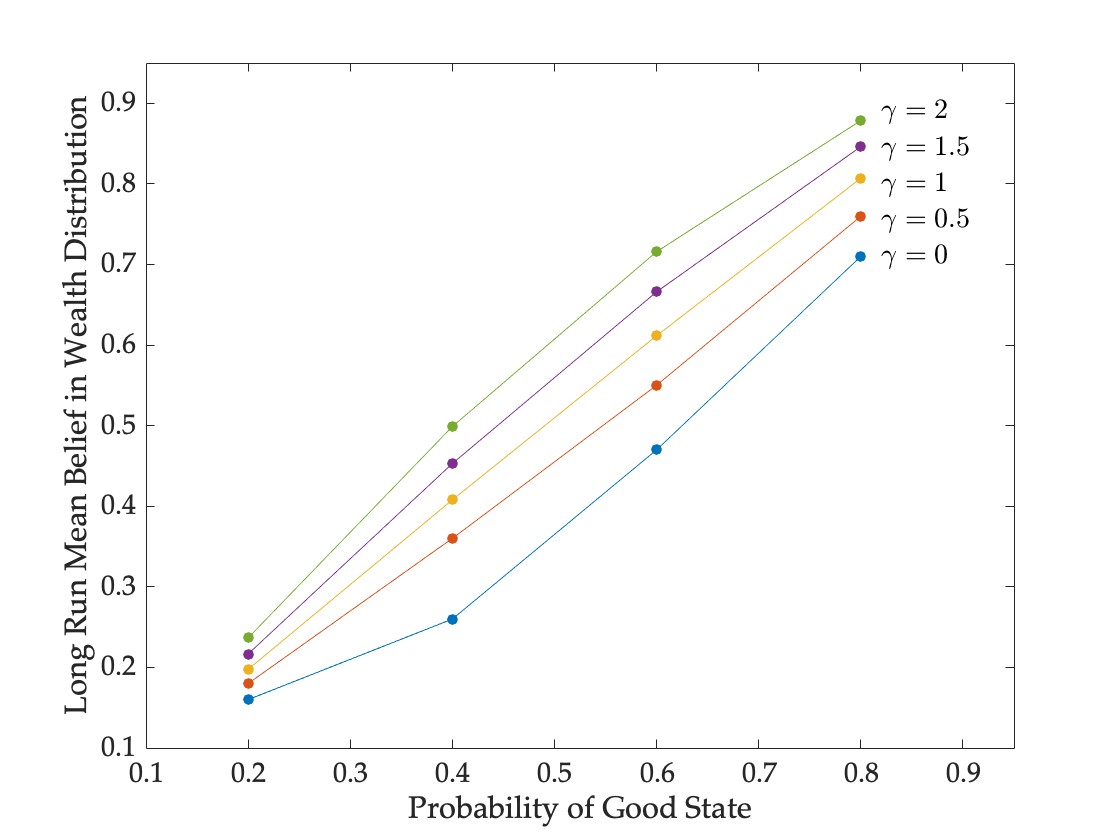}
\end{center}
\vspace{-0.25in}
\caption{Mean Beliefs in the long run for various values of $\gamma$ and $\pi^*$\label{f:robustness}}
\end{figure}

To explore the robustness of this finding, we simulate the model for a variety of risk aversion parameters and probabilities of a good state. Specifically, we consider $\gamma \in \{0, 0.5, 1, 1.5, 2 \}$ and $\pi^* \in \{0.2, 0.4, 0.6, 0.8 \}$. The mean of belief distribution of wealth after 5,000 periods is shown in Table \ref{t:robustness} and Figure \ref{f:robustness}.\footnote{Each cell in the table is based on the same random number seed, and hence for any given value of $\pi^*$, the same sequence of realized states. This filters out the effect of randomness from the comparison across different risk-preferences. Simulation results suggest that different random number seeds lead to different limiting belief distributions of wealth, while maintaining the qualitative properties revealed in the table. That is, the system is characterized by long run path dependence.}  Two patterns are apparent in the simulated data. First, for any (exogenously) given probability of a good state, greater risk aversion is associated with greater long run optimism in beliefs. And second, log utility results in the highest levels of accuracy in the long run. 

\begin{table}[h]
\centering
\begin{tabular}{c|cccc}
           & $\pi^* = 0.2$     & $\pi^* = 0.4$     & $\pi^* = 0.6$     & $\pi^* = 0.8$     \\ \hline 
  $\gamma = 0.0$   & 0.16 & 0.26    & 0.47    & 0.71 \\
        $\gamma = 0.5$ & 0.18  & 0.36 & 0.55 & 0.76 \\
      $\gamma = 1.0$   & 0.20 & 0.41 & 0.61 & 0.81 \\
        $\gamma = 1.5$ & 0.22 & 0.45 & 0.67 & 0.85 \\
        $\gamma = 2.0$   & 0.24 & 0.50 & 0.72 & 0.88 \\ \hline
\end{tabular}
\caption{Mean Beliefs in the long run for various values of $\gamma$ and $\pi^*$\label{t:robustness}}
\end{table}

We next relax the assumption that the the probability of default is exogenously given, and allow for the possibility that it depends on the degree of leverage.

\section{Endogenous Default Probabilities} \label{s:endogenous}

To this point we have considered the volume and terms of borrowing based on the distribution of wealth among investors with heterogeneous beliefs, treating the objective probability of a good state as exogenous. But the economy is a self-referential system---the objective probability of any state itself depends on the distribution of subjective beliefs. In this section we allow for the possibility that the likelihood of good and bad states (and hence the probability of default) depends on the scale of borrowing relative to total wealth. That is, we allow for the function $\pi(b)$ to be non-degenerate. 

\begin{figure}[t]
\begin{center}
\includegraphics[width=5in]{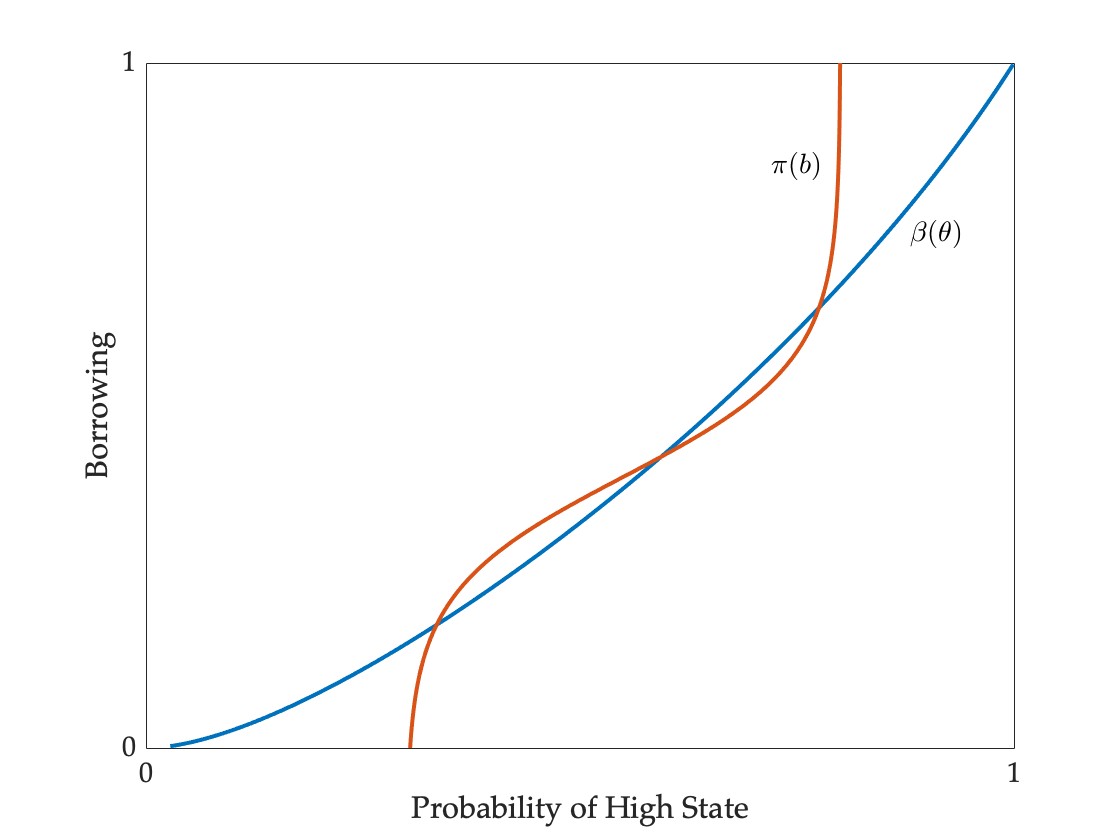}
\end{center}
\par
\vspace{-0.25in}
\caption{Borrowing given beliefs and probability of full repayment given borrowing \label{f:REequil}}
\end{figure}

In this case one case ask what values of $\theta$ will be self-fulfilling in the following sense. Suppose all wealth were held by belief type $\theta_i$ in period $t$. Then the share in bonds $\sigma_{it}$ of this belief type would equal the overall level of borrowing $b_t$, and the market clearing condition (\ref{market_clear}) would be redundant. Equilibrium values of $(b_t, p_t, q_t)$ would be given by (\ref{issue}--\ref{price}) and the solution to the maximization problem (\ref{e:objective}).

Let $\beta(\theta)$ denote the value of $b$ that arises in equilibrium when all wealth is held by belief type $\theta$. It is easily verified that this function must be increasing, given concavity of $h$ and $l$. For the case of log utility, the function $\beta(\theta)$ may be obtained from the solution (\ref{e:logsolution}). In particular, when the solution is interior, $\beta(\theta)$ is the unique value of borrowing $b$ that satisfies 
\[
\frac{\theta}{1-l(b)/b}-\frac{1-\theta}{h(b)/b-1} = b
\]
A particular case of this function is shown in Figure \ref{f:REequil}, using the same specifications for $h$ and $l$ as in the example in Section \ref{s:riskneutral}.

Also shown in Figure \ref{f:REequil} is one possible specification for the function $\pi(b)$, which identifies the objective probability of a high state given the extent of borrowing.\footnote{The figure is based on $\pi(b) = 0.3 + 0.5/(1+\exp(4.75-12b))$, for which the probability of a good state lies in the interval $(0.3, 0.8)$.} This would be a vertical line if the objective probability distribution were exogenous. In the example shown, $\pi(b)$ is an increasing function: higher levels of borrowing and expenditure result in increased economic activity, making default less likely, though still possible.\footnote{There are several ways in which such a relationship can be microfounded, starting with \citet{diamond1982aggregate}, with other approaches discussed in \citet{cooper1988coordinating} and \citet{aoki2007reconstructing} for example.} 

An intersection of the two curves in Figure \ref{f:REequil} can be viewed as a rational expectations equilibrium: if all agents held the corresponding belief, they would borrow at levels that give rise to an objective probability that matches the belief. In the example shown in the figure, there are multiple beliefs about the probability of default that are self-fulfilling. This suggests the possibility that the economy can alternate between high and low leverage regimes.

\begin{figure}[t]
\begin{center}
\includegraphics[width=5in]{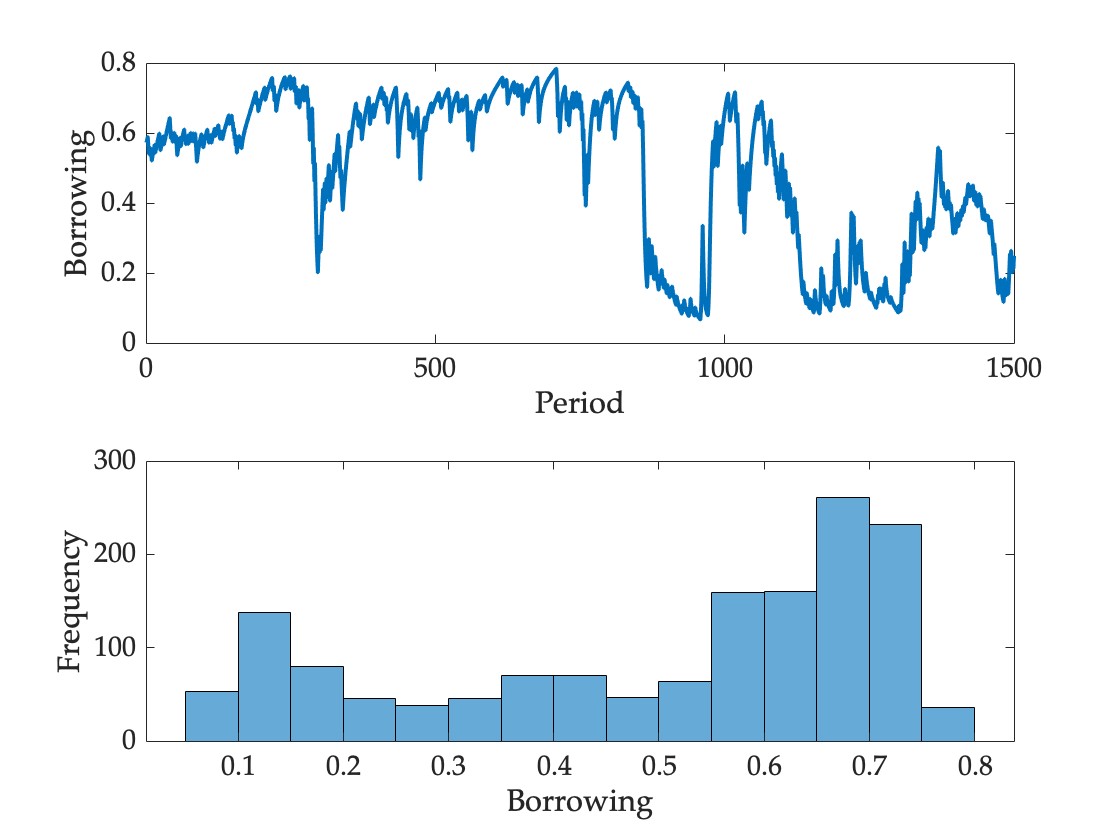}
\end{center}
\par
\vspace{-0.25in}
\caption{A Sample Path for the Dynamics of Leverage \label{f:leverage}}
\end{figure}

To explore this conjecture, we simulate the model for 2,500 periods, and add some noise to ensure that the wealth distribution does not simply converge to a single belief. Specifically, in each period, we take a weighted average of the inherited belief distribution of wealth $f_t(\theta)$ and a uniform belief distribution of wealth, with weight 0.01 on the latter. The last 1500 periods of the resulting time series are shown in the top panel of Figure \ref{f:leverage}. The bottom panel of Figure \ref{f:leverage} shows the distribution of leverage across the last 1,500 periods of the simulated time series. We see that the economy spends considerable periods of time in states with low levels of leverage, high notional interest rates (low bond prices), and a  high likelihood of default.  The wealth distribution of beliefs in such periods is concentrated among pessimists. A sequence of positive shocks can switch the economy to a more favorable regime, in which borrowing is high, notional interest rates are low, default probabilities are low, and wealth is concentrated among optimists. Transitions between these regimes occur endogenously and erratically.

\section{Conclusions}

We have explored a relatively simple model with heterogeneous prior beliefs about a binary state that governs aggregate outcomes. The key driver of the dynamics is the belief distribution of wealth, which determines the scale and terms of borrowing in equilibrium. This belief itself evolves as uncertainty is resolved, in favor of optimists if loans are repaid in full, and in favor of pessimists if there is default. An economy with these features traverses a path that involves periods of high leverage and low default probabilities, punctuated by periods of stagnation. 

There are a number of directions in which this work could be extended. We have not considered learning, for example. Individuals can learn in two ways---from observations of data or from observations of others. In this model those who try to learn from data can indeed come to have accurate beliefs in the long run, but this is not profitable, since those with such beliefs end up with diminishing relative wealth over time. If people learn from others, by adopting beliefs and behaviors that are successful, the resulting wealth dynamics would be very similar to those described here. 

We have also considered only one period loans, and it would be worth considering debt with different maturities, equity financing, and beliefs over different horizons. This would allow for an examination of liquidity in addition to leverage.

Given the abstract nature of the analysis here, we have refrained from making explicit policy recommendations. But an important policy concern is whether or not tools exist for fiscal and monetary authorities to shift a stagnating economy towards a regime with higher levels of economic activity. There is no obvious connection between traditional policy tools and the belief distribution of wealth, though intervention in the face of large scale default could have an effect. 

Our analysis here has allowed for heterogenous beliefs but has not explored the implications of heterogeneous risk preferences. An alternative approach could reverse this, by assuming heterogeneous risk preferences but common beliefs. In this case one would speak of the distribution of wealth across preference (rather than belief) types. More generally, one could allow for heterogeneity in both beliefs and preferences and consider the distribution of wealth across a set of multidimensional characteristics. This seems to be a promising generalization that we leave to future research.

\newpage

\appendix
\section*{Appendix}

\proof[Proof of Proposition \ref{p:unique}] We drop time subscripts for clarity, since these play no role in the within-period analysis. For each $\theta_i \in \Theta$ define a level of borrowing $\alpha_i$ such that belief type $i$ is indifferent between cash and bonds. From (\ref{indiff}), this may be obtained by solving
\begin{equation*} 
{\theta}_i \frac{h(\alpha_i)}{\alpha_i} + (1-{\theta}_i) \frac{l(\alpha_i)}{\alpha_i} =  1.
\end{equation*}
Since $l$ is concave and $h$ is strictly concave, the solution is unique, and satisfies $0 = \alpha_1 < ... < \alpha_n = 1$. Define $\omega_i = 1 - F(\theta_i) + f(\theta_i)$ as the share of wealth held by belief types who are at least as optimistic as $\theta_i$. Then we have $1 = \omega_1 > ... > \omega_n = f(\theta_n)$. Since $\alpha_1 < \omega_1$ and $\alpha_n > \omega_n$, there must exist some belief type $\theta_i$ such that $\alpha_i < \omega_i$ and $\alpha_{i+1} \ge \omega_{i+1}$. 

We consider two cases. 

If $\alpha_i \le \omega_{i+1}$, there is a unique equilibrium $b = \omega_{i+1}$ in which beliefs types $\theta_{i+1}, ... , \theta_n$ invest fully in bonds, while the rest remain fully in cash. Even though type $\theta_i$ may be indifferent at this equilibrium, this type would have a strict preference for cash if even a small share of those with belief $\theta_i$ were to invest in bonds. 

If $\alpha_i > \omega_{i+1}$, there is a unique equilibrium $b = \omega_{i+1} + s_i f(\theta_i)$ in which beliefs types $\theta_{i+1}, ... , \theta_n$ invest fully in bonds along with a share $s_i \in (0,1)$ of type $\theta_i$, while the rest remain in cash. The critical type is indifferent while the rest strictly prefer to invest or abstain. 

Hence the scale of borrowing $b$ and the critical type $\bar{\theta}$ are pinned down by the belief distribution of wealth. This determines the size of the bond issue by (\ref{issue}) and the bond price by (\ref{price}). \endproof

\proof[Proof of Proposition \ref{p:compstat}] We prove the result for a positive shock, the argument
for a negative shock is analogous. As in the proof of Proposition 1, define $\alpha_i$ as the level of borrowing at which type $\theta_i$ would be indifferent between borrowing and lending based on (\ref{indiff}), and $\omega_i$ as the share of wealth held by type $\theta_i$ together with all those who are more optimistic, based on the distribution $f_t(\theta)$. In addition, define $\omega'_i$ as the share of wealth held by type $\theta_i$ together with all those who are more optimistic, based on the distribution $f_{t+1}(\theta)$. Following a positive shock, $\omega'_i > \omega_i$ for all $i > 1$. Recall (from the proof of Proposition 1)  that types with $\alpha_i \le \omega_{i+1}$ do not invest in period $t$. Since $\alpha_i \le \omega_{i+1}$ implies $\alpha_i \le \omega'_{i+1}$ after a positive shock, such types also do not invest in period $t+1$. Hence $\bar{\theta}_{t+1} \ge \bar{\theta}_t$. 

Next consider the volume of lending. Suppose, by way of contradiction, that $b_{t+1} < b_t$ following a positive shock. Then, from (\ref{indiff}), any type that was indifferent at $t$ will strictly prefer to lend at $t+1$, and all types that strictly preferred to lend will continue to do so. Given that the share of the aggregate wealth of these types has increased, we must have $b_{t+1} \ge b_t$, a contradiction. Hence $b_{t+1} \ge b_t$. This immediately implies $h(b_{t+1}) \ge h(b_t)$ and hence $q_{t+1} \ge q_t$ from (\ref{issue}). And since $p_{t+1} = b_{t+1}/h(b_{t+1})$ from (\ref{price}), concavity of $h$ implies $p_{t+1} \ge p_t$. \endproof

\proof[Proof of Proposition \ref{p:unique-log}] We consider $b_{t}-\sum_{\theta_i}\sigma_{it} f_{t}(\theta_i)$
as a function of $b_{t}$ and will show that there is a unique value
of $b_{t}$ for which 
$$b_{t}-\sum_{\theta_i \in \Theta}\sigma_{it} f_{t}(\theta_i)=0,$$ 
as required by the market clearing condition (\ref{market_clear}). This establishes uniqueness of $b_{t}$. Uniqueness of $q_{t}$ then
follows from $q_{t}=h(b_{t})$, and $p_{t}$ is
uniquely determined by $p_{t}=b_{t}/h(b_{t})$. 

Using (\ref{issue}) and (\ref{price}), we can rewrite the optimal solution
to the portfolio choice of the investor with belief type $\theta$
in terms of $b_{t}$ as follows:
\[
\sigma_{it}=\begin{cases}
0 & \textrm{if}\:\theta\leq\theta_{t}^{\min,b}\\
\frac{\theta}{L_{t}^{b}}-\frac{1-\theta}{H_{t}^{b}} & \textrm{if}\:\theta_{t}^{\min,b}<\theta_{i}\leq\theta_{t}^{\max,b}\\
1 & \textrm{otherwise}
\end{cases},
\]
where 
\[
L_{t}^{b}\equiv1-\frac{l\left(b_{t}\right)}{b_{t}},\;H_{t}^{b}\equiv\frac{h\left(b_{t}\right)}{b_{t}}-1,
\]
and 
\[
\theta_{t}^{\min,b}\equiv\frac{L_{t}^{b}}{L_{t}^{b}+H_{t}^{b}},\;\theta_{t}^{\max,b}\equiv\frac{H_{t}^{b}L_{t}^{b}+L_{t}^{b}}{H_{t}^{b}+L_{t}^{b}}.
\]
Note that $\sigma_{it}$ is a continuous function of $b_{t}.$ We
also observe that $\theta_{t}^{\min,b}$ and $\theta_{t}^{\max,b}$
are strictly increasing functions of $b_{t}$ and $\theta/L_{t}^{b}-\left(1-\theta\right)/H_{t}^{b}$
is a strictly decreasing function of $b_{t}$. Further, $0<\theta_{t}^{\min,b}<\theta_{t}^{\max,b}<1$
for $b_{t}\in(0,1).$ This indicates that $\sigma_{it}$ is a decreasing
function of $b_{t}$. Under Assumption 1 , at $b_{t}=1,$ we have
$\theta_{t}^{\min,b}=1$, hence $\sigma_{it}=0$ for all $\theta$
and thus $b_{t}-\sum_{\theta_{i}}f_{t}(\theta_{i})\sigma_{it}=1>0$.
For $b_{t}\downarrow0,$ we $\theta_{t}^{\min,b}\downarrow0$ and
$\theta_{t}^{\max,b}=1-\lim_{b\downarrow0}l(b)/b<1$. Hence there
exists some $\theta$ for which $\sigma_{it}=1$ as $b_{t}\downarrow0$.
Thus $b_{t}-\sum_{\theta_{i}}f_{t}(\theta_{i})\sigma_{it}$ is negative
for $b_{t}\downarrow0$. Since $\sigma_{it}$ is a continuous function
of $b_{t}$, $b_{t}-\sum_{\theta_{i}}f_{t}(\theta_{i})\sigma_{it}$
is also a continuous function of $b_{t}$. Further $b_{t}-\sum_{\theta_{i}}f_{t}(\theta_{i})\sigma_{it}$
is a strictly increasing function of $b_{t}$ for $b_{t}\in(0,1)$.
This indicates that there is a unique value of $b_{t}$ for which
$b_{t}-\sum_{\theta_{i}}f_{t}(\theta_{i})\sigma_{it}=0$. 

Next, we prove the comparative static
result for a positive shock, the argument for a negative shock is
analogous. Observe that $\sigma_{it}$ is a weakly increasing function
of $\theta$, so that investors with higher beliefs about the likelihood
of a positive shock invest a (weakly) higher fraction of their wealth
in bonds. The positive shock thus leads to $F_{t+1}(\theta)\leq F_{t}(\theta)$,
that is the relative share of wealth has increased for those who invested
more. Suppose by way of contradiction, that $b_{t+1}<b_{t}$ following
a positive shock. Note that in the market clearing condition, the
supply of loans is a weighted average of share of wealth invested
by different investors, with the weights being equal to the investor's
share of aggregate wealth. As a result of the positive shock, the
share of aggregate wealth has shifted towards those who invested more.
Since $\sigma_{it}$ is a weakly decreasing function of $b$, we have
that everyone (weakly) prefers to lend more. Hence we must have $b_{t+1}\geq b_{t},$
a contradiction. Hence $b_{t+1}\ge b_{t}$. This immediately implies
$h(b_{t+1})\ge h(b_{t})$ and hence $q_{t+1}\ge q_{t}$ from (\ref{issue}).
And since $p_{t+1}=b_{t+1}/h(b_{t+1})$ from (\ref{price}), concavity
of $h$ implies $p_{t+1}\ge p_{t}$. \endproof

\newpage

\bibliographystyle{plainnat}
\bibliography{macro}

\end{document}